
\documentstyle{amsppt}
\TagsOnRight
\catcode`\@=11
\def\logo@{}
\catcode`\@=13
\parindent=8 mm
\magnification 1200
\hsize = 6. true in
\vsize = 8.5 true in
\hoffset = .4 true in
\parskip=\medskipamount
\baselineskip=14pt
\def \oc {\overset {\otimes}\to{,}}
\def \di{\partial}

\def \smaller {\eightpoint}
\def \wt {\widetilde}
\def \wh {\widehat}

\def \ra {\rightarrow}

\def \lra {\longrightarrow}
\def \lmt {\longmapsto}

\def \d {\delta}

\def \k {\kappa}

\def \l {\lambda}

\def \r {\rho}

\def \o {\omega}

\def \ss {\subset}

\def \DD {\Cal D}

\def \HH {\Cal H}

\def \NN {\Cal N}

\def \di {\partial}

\def \tr{\text{tr}}

\def\rk{\text{\rm rk}}

\def \End{\text{End}}

\redefine\cite#1{{\bf[#1]}}
\phantom{0}\vskip -.5 true in
\noindent hep-th/9406078  \hfill  CRM-2890 (1994)
\vskip .5 true in
\topmatter
\title
Quantum Isomonodromic Deformations and the Knizhnik--Zamolodchikov
Equations
\endtitle
\leftheadtext{J\. Harnad}
\rightheadtext{Quantum Isomonodromic Deformations}
\author
J\. Harnad
\endauthor
\affil
Department of Mathematics and Statistics, Concordia
University \\
7141 Sherbrooke W., Montr\'eal, Canada H4B 1R6, and \\
Centre de recherches math\'ematiques, Universit\'e de Montr\'eal \\
 C\. P\. 6128, Succ. centre--ville, Montr\'eal, Canada H3C 3J7 \\
e-mail: {\it harnad\@alcor.concordia.ca}  or  {\it harnad\@mathcn.umontreal.ca}
\endaffil
\abstract
Viewing the Knizhnik--Zamolodchikov equations as multi--time, nonautonomous
Shr\"odinger equations, the transformation to the Heisenberg representation is
shown to yield the quantum Schlesinger equations. These are the quantum form of
the isomonodromic deformations equations for first order operators of the form
$\DD_\l = {\di \over \di \l} - \wh{\NN}(\l)$, where $\wh{\NN}(\l)$ is a
rational $r\times r$ matrix valued function of $\l$ having simple poles only,
and the matrix entries are interpreted as operators on a module of the rational
$R$--matrix loop algebra $\wt{\frak{gl}}(r)_R$. This provides a simpler
formulation
of a construction due to Reshetikhin, relating the KZ equations to quantum
isomonodromic deformations.
\endabstract
\thanks Research supported in part by the
Natural Sciences and Engineering Research Council of Canada and the Fonds FCAR
du Qu\'ebec.
\endthanks
\endtopmatter
\document
\baselineskip=14pt
\subheading{1. Schlesinger Equations as Nonautonomous Hamiltonian Systems}
\nopagebreak \medskip
   In \cite{H, HTW}, it was shown how the linear  rational $R$--matrix
structure on the loop algebra $\wt{\frak{gl}}(r)_R$, when combined with a
parametric family of Poisson embeddings into the dual space
$\wt{\frak{gl}}(r)^*_R$, gives rise to the Schlesinger equations
$$
\align
{\di N_i\over \di z_j} = & \phantom{-}{[N_i, N_j]\over z_i - z_j}, \quad i\neq
j, \quad i,j =1, \dots , n  \tag{1.1a}
\\
{\di N_i \over \di z_i} =& - \sum_{j=1\atop j\neq i}^n
{[N_i,\ N_j]\over z_i - z_j}.  \tag{1.1b}
\endalign
$$
Here $\{N_i\}_{i=1,\dots n}$ is a set of $r\times r$ matrices depending on the
$n$ complex parameters $\{z_i\}_{i=1, \dots n}$. Eqs\. (1.1a,b) may be viewed
as
deformation equations that preserve the monodromy of the differential operator
$$
\DD_\l := {\di\over \di \l} - \NN(\l),  \tag{1.2}
$$
where
$$
\NN(\l):= \sum_{i=1}^n {N_i \over \l - z_i}, \tag{1.3}
$$
about the regular singular points $\{z_1, \dots z_n, \infty\}$. They may also
be
viewed as a compatible set of nonautonomous Hamiltonian equations
induced by the Poisson commuting set of Hamiltonians
$$
H_i = \sum_{j=1\atop j\neq i}^n {\tr(N_i N_j)\over z_j -z_i}
\tag{1.4}
$$
under the Lie--Poisson bracket structure determined by
$$
\{(N_i)_{ab},\ (N_j)_{cd}\} = \d_{ij} \left((N_i)_{ad}\d_{cb} -
(N_i)_{cb}\d_{ad}
\right).
\tag{1.5}
$$
When viewed as autonomous systems, the Hamiltonians (1.4) determine the
classical
limit of the Gaudin spin chain models \cite{G1, G2}.  Here, however,  the time
parameters are identified with the deformation parameters $\{z_1, \dots
,z_n\}$,
and hence enter explicitly in the Hamiltonians, making the resulting systems
multi--time nonautonomous systems.

   Eqs\. (1.1a,b) may equivalently be expressed as the commutativity conditions
$$
[\DD_\l,\ \DD_i]=0  \tag{1.6}
$$
between $\DD_\l$ and the infinitesimal deformation operators
$$
\DD_i := {\di \over \di z_i} + {N_i \over \l -z_i},  \tag{1.7}
$$
from which the invariance of the monodromy of the operator
$\DD_\l$ around the singular points may be deduced \cite{JMMS, JMU}. The
compatibility conditions
$$
[\DD_i,\ \DD_j] = 0  \tag{1.8}
$$
follow from the Poisson commutativity
$$
\{H_i,\ H_j\}=0  \tag{1.9}
$$
of the Hamiltonians with respect to the classical $R$--matrix Poisson bracket
structure \cite{FT, ST}
$$
\{\NN(\l)\oc N(\mu)\} = [r(\l-\mu), \NN(\l)\otimes I +
I \otimes \NN(\mu)],  \tag{1.10}
$$
where
$$
r(\l) := {P_{12}\over \l},  \tag{1.11}
$$
with $P_{12}$ the permutation operator on $\bold{C}^r\otimes \bold{C}^r$. The
involutivity conditions (1.9) follow from the fact that the $H_i$'s are just
the residues at $\l=z_i$ of the rational spectral invariant function
$$
\align
\Delta(\l) = & -{1\over 2} \NN^2(\l)  \\
   =& \sum_{i=1}^n {H_i\over \l-z_i} -{1\over 2} \sum_{i=1}^n {\tr (N_i^2)
\over
(\l-z_i)^2},
\tag{1.12}
\endalign
$$
which, because of the $R$--matrix structure (1.10) satisfies
$$
\{\Delta(\l),\ \Delta(\mu)\} =0 \quad  \forall\ \l, \mu \in \bold{C}.
\tag{1.13}
$$

   As shown in \cite{H}, the proper interpretation of the system (1.1a,b) as a
nonautonomous system of compatible Hamiltonian equations is not in the space
$\wt{\frak{gl}}(r)^*_R$, but in an auxiliary symplectic space $M$, consisting
of
pairs $(F,G)$ of $N\times r$ rectangular matrices, where
$$
N=\sum_{i=1}^n k_i,  \tag{1.14}
$$
and $k_i=\rk(N_i)$, with canonical symplectic form
$$
\o := \tr(dF^T\wedge dG).  \tag{1.15}
$$

The relation between the two spaces is given by the parametric family of
Poisson
maps \cite{AHH, H, HW1}
$$
\align
J_A:M &\lra \wt{\frak{gl}}(r)^*_R  \tag{1.16a}\\
J_A:(F,G) &\lmt -G^T(A-\l I)^{-1} F :=\NN(\l),  \tag{1.16b}
\endalign
$$
where $A$ is a diagonal $N\times N$ matrix with eigenvalues
$\{z_i\}_{i=1\dots n}$ of multiplicities $\{k_i\}_{i=1\dots n}$. (Note that the
$R$--matrix Poisson bracket relations (1.10) remain valid if $A$ is replaced by
an
arbitrary $N\times N$ matrix; cf\. \cite{HW1}). The Hamiltonian equations on
$M$
generated by the Hamiltonian
$$
\wt{H}_i = H_i \circ J_a, \tag{1.17}
$$
with the identification of the  time variables  again with $\{z_1, \dots ,
z_n\}$,
are then
$$
\align
{\di F_i \over \di z_j} =&\phantom{-} F_i {G_j^T F_j\over z_i-z_j},  \qquad
i\neq
j  \tag{1.18a}\\
{\di F_i \over \di z_i}= & -\sum_{j=1\atop j\neq i}^n F_i {G_j^T F_j\over
z_i-z_j},  \tag{1.18b} \\
{\di G_i \over \di z_j} =& - G_i {F_j^T G_j\over z_i-z_j},
\qquad i\neq
j  \tag{1.18c}\\
{\di G_i \over \di z_i} =&\phantom{-}\sum_{j=1\atop j\neq i}^n G_i
{F_j^T G_j\over z_i-z_j},
\tag{1.18d}
\endalign
$$
where $(F_i, G_i)$ denote the $i$th  blocks in $(F,G)$, of dimension $k_i
\times
r$, corresponding to the eigenvalue $z_i$. Eqs\. (1.18a-d), combined with the
parameter--dependent Poisson maps $J_A$ then imply the isomonodomic deformation
equations (1.1a,b).

   In the following section, it will be shown how the quantum analogue of the
above construction gives rise, within the Shr\"odinger representation, to the
Knizhnik-Zamolodchikov equations \cite{KZ} determining the $n$--point
correlation functions in the WZWN model, while in the Heisenberg
representation, they give the quantum  version of the Schlesinger equations
(1.1a,b). This formulation was suggested by the work of Babujian and Flume
\cite{B, BF} on the relation between the Knizhnik--Zamolodchikov equations
and the Bethe ansatz method for Gaudin spin chains. A related, though
somewhat more complicated formulation of the link between the KZ equations and
quantum isomonodromic deformations has been given by Reshetikhin \cite{R}.
The present version corresponds to a choice of complex, simple Lie algebra
$\frak{g}=\frak{sl}(r,\bold{C})$ and simple poles at $\{\l=z_i\}$. The case of
higher order poles may be similarly dealt with by choosing the matrix $A$ in
eq\. (1.16b) to have nondiagonal Jordan structure (cf\. \cite{AHP, HW1, HW2}).
  \bigskip
\pagebreak
\subheading{2. Quantum Schlesinger System and the Knizhnik--Zamolodchikov
Equations}
\nopagebreak\medskip
   Let $\frak{g}\ss\frak{gl}(r, \bold{C})$ be a matrix Lie algebra and
$\{\HH_i\}_{i=1, \dots n}$ a set of $\frak{g}$--modules on which the
representations $\r_i\ra \End(\HH_i)$ are defined. In the following, $\frak{g}$
will just be taken as $\frak{gl}(r, \bold{C})$, but subalgebras and real forms
may easily be similarly dealt with. Let
$$
\HH:= \otimes_{i=1}^n \HH_i  \tag{2.1}
$$
be the tensor product space, and denote by
$$
\align
\tilde{\r}_i: \frak{g} &\lra \End(\HH)\\
\tilde{\r_i}: X &\lmt I\otimes \cdots \underbrace{\r_i(X)}_{i\text{th factor}}
\cdots \otimes I, \qquad X\in \frak{g} \tag{2.2}
\endalign
$$
the extension of $\r_i$ to $\HH$. Let $\wh{N}_i$ be the $\End(\HH)$--valued
$r\times r$ matrix with elements
$$
(\wh{N}_i)_{ab} =\wh{\r}_i(E_{ab}),  \tag{2.3}
$$
where $\{E_{ab}\}$ is the standard basis for $\frak{gl}(r)$ consisting of the
elementary matrices with nonvanishing entries in the
$(ab)$th position. These then satisfy the $\frak{gl}(r)$ commutation relations
$$
[(\wt{N}_i)_{ab} ,\ (\wt{N}_j)_{cd}] =
\d_{ij} \left( (\wt{N}_i)_{ad} \d_{bc} - (\wt{N}_i)_{bc} \d_{ad}\right).
\tag{2.4}
$$
Defining the $\frak{gl}(r) \otimes \End(\HH)$--valued rational function
$$
\wt{\NN}(\l) := \sum_{i=1}^n {\wt{N}_i\over \l - z_i},  \tag{2.5}
$$
we see that this defines a representation of the $\wt{gl}(r)_R$ linear
$R$--matrix
Lie algebra
$$
[\wt{\NN}(\l),\ \wt{\NN}(\mu)] =
[r(\l-\mu),\ \wt{\NN}(\l) \otimes I + I \otimes  \wt{\NN}(\mu)].  \tag{2.6}
$$

   For example, $\HH$ could be taken as the space of polynomials in the
matrix elements of the $N\times r$ matrix $G$, with each factor $\HH_i$
corresponding to polynomials in the block $G_i$ only, and $\wt{\NN}(\l)$
defined by replacing $F$  in eq\. (1.16b) by the gradient operator $F\ra
\nabla_G$. This then implies the commutation relations (2.4) and (2.6) in the
case of diagonal $A$. If we choose maximal rank $k_i =r$ for each eigenvalue
block
$G_i$, a suitable completion of $\HH_i$ may be identified with the space of
differentiable functions on the group $\frak{Gl}(r)$, and the resulting
representation is simply given by the left--invariant vector fields.
Alternatively,
for rank $k_i < r$, we may view the block $G_i$ as homogeneous coordinates on
the
complex Grassmannian $\frak{Gr}_{k_i}(\bold{C}^r)$, and pass to the quotient
space
under the natural action of $\frak{Gl}(k_i)$ on $k_i$--frames in $\bold{C}^r$.
The
resulting representation then consists of vector fields
on $\frak{Gr}_{k_i}(\bold{C}^r)$ induced by the infinitesimal
$\frak{Gl}(r)$--action. More generally,  replacing $A$ by an arbitrary $N\times
N$
matrix in complex Jordan normal form, with eigenvalues $\{z_i\}$, the
commutation
relations (2.6) still hold, but the form of (2.5) changes to a matrix--operator
valued rational function of $\l$ with poles at each of the eigenvalues $\{z_1,
\dots , z_n\}$ of order equal to the dimension of the largest corresponding
Jordan
block. The resulting algebra associated to each $z_i$ is no longer
$\frak{gl}(r)$,
but the jet extension $\frak{gl}(r)^{(l_i)}$, where $l_i+1$ is the order of the
pole
at $z_i$ (cf. \cite{R}). In the following, we remain with the case of simple
poles
only.

   Defining the $\End(\HH)$--valued rational function
$$
\wt{\Delta}(\l) := {1\over 2} \tr(\wt{\NN}(\l))^2
= \sum_{i=1}^n{\wt{H}_i\over\l - z_i} +
\sum_{i=1}^n{\tr(\wt{N}_i^2)\over(\l -z_i)^2},  \tag{2.7}
$$
where
$$
\wt{H}_i =  \sum_{j=1 \atop j \neq i}^n
{\tr(\wt{N}_i \wt{N}_j)\over z_i - z_j},  \tag{2.8}
$$
it follows from (2.6), just as in the classical case, that
$$
[\wt{\Delta}(\l), \ \wt{\Delta}(\mu)] =0, \quad \forall\ \l, \mu \in \bold{C}
\tag{2.9}
$$
and hence
$$
[\wt{H}_i, \ \wt{H}_j] =0,  \quad   i,j=1, \dots n.   \tag{2.10}
$$

  The stationary states of the $n$--site $\frak{gl}(r)$ Gaudin spin chain are
the simultaneous eigenvectors of the operators $\{\wt{H}_i\}_{i=1 \dots n}$.
These may, in principle, be constructed via the algebraic Bethe ansatz
\cite{J, FFR}. However, instead of considering stationary states, we consider
the
time--dependent Shr\"odinger equations
$$
i\hbar {\di \Psi \over \di t_i} = \wt{H}_i\Psi, \quad \Psi \in \HH.  \tag{2.11}
$$
These equations are closely related to the Knizhnik-Zamolodchikov equations
determing the $n$--point correlation functions for the WZWN model
$$
\k {\di \Psi \over \di z_i} = \wt{H}_i\Psi,
  \tag{2.12}
$$
where
$$
\k = {1\over 2} (c_\HH + c_{\frak{g}}),  \tag{2.13}
$$
$c_\HH$ being the level and $c_{\frak{g}}$ the dual Coxeter number. Eq\. (2.12)
is
obtained from (2.11) by identifying, as in the classical case, $t_i \sim z_i$
and
choosing ${\k \over i\hbar}\wh{H}_i$ as Hamiltonian. This allows us to view
(2.12)
as a system of compatible, multi-time nonautonomous Shr\"odinger equations.

   We may pass, as usual, from the Shr\"odinger to the Heisenberg
representation
by introducing the (multi--time) ``evolution'' operator $U(z_1, \dots z_n) \in
\End(\HH)$, which is the unique invertible operator satisfying the  equation
$$
\k{\di U \over \di z_i} =  U \wt{H}_i  \tag{2.14a}
$$
with initial conditions
$$
U(z_1^0, \dots, z_n^0) = Id.  \tag{2.14b}
$$
The operator--valued matrices $\wt{N}_i$ in the Heisenberg representation then
become
$$
\wh{N}_i := U \wt{N}_i U^{-1},  \tag{2.15}
$$
where the conjugation by $U$ is understood as applied to each matrix element in
$\wt{N}_i$. Correspondingly, we have the Heisenberg representation of the
rational
$\frak{gl}(r)\otimes \End(\HH)$--valued function $\wt{\NN}(\l)$
$$
\wh{\NN}(\l) :=U\wh{\NN}(\l)U^{-1} = \sum_{i=1}^n {\wh{N}_i \over \l -
z_i},  \tag{2.16}
$$
and of the Hamiltonians
$$
\wh{H}_i =U\wt{H}_i U^{-1}.  \tag{2.17}
$$
Differentiating the $\wh{N}_i$'s with respect to the $z_j$'s, and using
(2.14a),
gives
$$
\align
\k {\di \wh{N}_i \over \di z_j} =&\phantom{-} [\wh{H}_j,\ \wh{N}_i]
 = {[\wh{N}_i, \ \wh{N}_j] \over z_i - z_j},  \qquad i\neq j   \tag{2.18a}\\
\k {\di \wh{N}_i \over \di z_i}
=& -\sum_{j=1\atop j\neq i}^n[\wh{H}_j,\ \wh{N}_i]
 =  -\sum_{j=1\atop j\neq i}^n{[\wh{N}_i, \ \wh{N}_j] \over z_i - z_j},
 \tag{2.18b}
\endalign
$$
where the second equality in (2.18a,b) follows from eqs\. (2.5), (2.7).
Absorbing
the factor $\k$ into the definition of $\wt{N}_i$, eqs\. (2.18a,b) are just the
quantum version of the Schlesinger equations (1.1a,b). Defining the
$\End(\HH)$--valued matrix differential operators
$$
\align
\wh{\DD}_\l :=& {\di \over \di \l} -
{1\over \k} \sum_{i=1}^n{\wh{N}_i\over \l-z_i}  \tag{2.19a}\\
\wh{\DD}_i :=& {\di \over \di z_i} + {1\over \k} {\wh{N}_i\over \l - z_i},
\tag{2.19b}
\endalign
$$
eqs\. (2.18a,b) are equivalent to the ``quantum isomonodromic deformation''
equations
$$
[\wh{\DD}_\l, \ \wh{\DD}_i] = 0.  \tag{2.20}
$$
The compatibility conditions
$$
[\wh{\DD}_i, \ \wh{\DD}_j] = 0   \tag{2.21}
$$
are again equivalent to the commutativity conditions (2.9) for the Hamiltonians
$\wt{H}_i$.

  The more elaborate construction of \cite{R} allows for higher order poles
in $\wh{\NN}(\l)$ and an arbitrary complex, simple Lie algebra $\frak{g}$. This
leads to a similar relation between the generalized rational KZ equations and
quantum isomonodromic deformation equations involving irregular singular points
at
$\{\l=z_i\}_{i=1, \dots n}$ (cf\. \cite{JMU, JM}).

 \bigskip \bigskip
\noindent{\it Acknowledgements.} The author would like to thank  A. Its for
helpful discussions and for bringing ref\. \cite{R} to his attention.
\bigskip\bigskip
\centerline{\bf References} \nobreak \bigskip  {\smaller
\item {\bf[AHH]}
 Adams, M\.R\., Harnad, J\. and  Hurtubise, J\.,
``Dual Moment Maps to Loop Algebras'', {\it Lett\. Math\. Phys\.} {\bf 20},
 294--308 (1990).
\item{\bf[B]}  Babujian, H\.M\. ``Off--Shell Bethe Ansatz Equations and
$N$--Point Correlators in the $SU(2)$ WZNW Theory'', {\it J\. Phys\.} {\bf A
26}
6981--6990 (1993).
\item{\bf[BF]}  Babujian, H\.M\. and Flume, R\., ``Off--Shell Bethe Ansatz
Equation for Gaudin Magnets and Solutions of Knizhnik--Zamolodchikov
Equations'', preprint, Bonn (1993)
\item{\bf[FT]} Faddeev, L\.D\. and Takhtajan, L.A.,
{\sl Hamiltonian Methods in the Theory of Solitons},\break
 Springer--Verlag, Heidelberg (1987).
\item{\bf[FFR]} Feigin, B\.,  Frenkel, E\. and  Reshetikhin, N\., ``Bethe
Ansatz
and Correlation Functions at the Critical Level'', preprint hep-th/9402022
(1994).
\item{\bf[G1]} Gaudin, M\., ``Diagonalization d'-une classe d'hamiltoniens de
spin'', {\it J\. Physique} {\bf 37}, 1087--1098 (1976).
\item{\bf[G2] }Gaudin, M\., {\sl La fonction d'onde de Bethe}, Masson, Paris
(1983).
\item{\bf[H]} Harnad, J\., ``Dual Isomonodromic Deformations and Moment Maps
into  Loop Algebras'', {\it Commun\. Math\. Phys\.} (1994, in press).
\item{\bf[HTW]} Harnad, J\., Tracy, C\.A\., Widom, H\., ``Hamiltonian Structure
of
Equations Appearing in Random Matrices'', in:  {\sl Low Dimensional Topology
and Quantum Field Theory}, ed. H. Osborn,  (Plenum, New York,  1993).
\item{\bf[HW1]} Harnad, J\.  and Wisse, M.--A\., ``Moment Maps to Loop
Algebras,
Classical $R$--Matrix  and Integrable Systems'', in: {\sl Quantum Groups,
Integrable Models  and Statistical Systems}  (Proceedings of the 1992 NSERC-CAP
Summer Institute in Theoretical Physics, Kingston, Canada, July 1992), ed.
J\. Letourneux and  L\. Vinet, World Scientific, Singapore (1993).
\item{\bf[HW2]} Harnad, J\. and Winternitz,  P\.,  ``Integrable Systems in
$\wt{\frak{gl}}(2)^{+*}$ and Separation of  Variables II. Generalized Quadric
Coordinates'', preprint CRM (1994).
\item{\bf [J]} Jur\v co, B\. ``Classical Yang--Baxter Equations and Quantum
Integrable Systems '', {\it J\. Math\. Phys\.} {\bf 30}, 1289--1295 (1989).
 \item{\bf[JMMS]} Jimbo, M\., Miwa, T\., M\^ori, Y\. and Sato, M\., ``Density
Matrix of an Impenetrable Bose Gas and the Fifth Painlev\'e Transcendent'',
{\it Physica} {\bf 1D}, 80--158 (1980).
\item{\bf[JMU]} Jimbo, M\., Miwa, T\., Ueno, K\., ``Monodromy Preserving
Deformation of Linear Ordinary Differential Equations with Rational
Coeefficients I.'', {\it Physica} {\bf 2D}, 306--352 (1981).
 \item{\bf[JM]} Jimbo, M\., Miwa, T\., ``Monodromy Preserving
Deformation of Linear Ordinary Differential Equations with Rational
Coeefficients II, III.'', {\it Physica} {\bf 2D}, 407--448 (1981); {\it ibid.},
{\bf 4D}, 26--46 (1981).
\item{\bf[KZ]}
Knizhnik, V\.G\. and Zamolodchikov, A\.B\.,``Current Algebra and Wess-Zumino
Model
in Two Dimensions'', {\it Nucl\. Phys\.} {\bf B 247},   83--103 (1984).
\item{\bf[R]} Reshetikhin, N\., ``The Knizhnik--Zamolodchikov System as a
Deformation of the Isomonodromy Problem'', {\it Lett\. Math\. Phys\.} {\bf 26},
167--177 (1992).
\item{\bf[ST]} Semenov-Tian-Shansky, M\.A\., ``What is a classical
$R$-matrix'', {\it Funct\. Anal\. Appl\.} {\bf 17} (1983) 259--272. \medskip}
\enddocument